\documentclass[a4paper]{article}

\usepackage{itgspeech2023}    
\usepackage{times}            
\usepackage[english]{babel}   
\usepackage[ansinew]{inputenc}
\usepackage[T1]{fontenc}      
\usepackage[sort&compress,numbers]{natbib}	
\usepackage{amsmath,amssymb}
\usepackage{graphicx}
\usepackage[colorlinks=false,pdfborder={0 0 0}]{hyperref}
\usepackage{units}

\usepackage{tikz}
\usepackage{booktabs}
\usepackage{microtype}
\usepackage[nolist, nohyperlinks]{acronym}

\newcommand\nnfootnote[1]{%
  \begin{NoHyper}
  \renewcommand\thefootnote{}\footnote{#1}%
  \addtocounter{footnote}{-1}%
  \end{NoHyper}
}

\title{Audio-Visual Speech Enhancement with Score-Based Generative Models}

\author{Julius Richter$^1$, Simone Frintrop$^2$, Timo Gerkmann$^1$}

\address{$^1$Signal Processing Group, $^2$Computer Vision Group, Department of Informatics, Universit\"at Hamburg, Germany\\
Email: \texttt{\{julius.richter,simone.frintrop,timo.gerkmann\}@uni-hamburg.de}}

\begin{document}

\begin{acronym}
\acro{sgm}[SGM]{score-based generative model}
\acro{snr}[SNR]{signal-to-noise ratio}
\acro{gan}[GAN]{generative adversarial network}
\acro{vae}[VAE]{variational autoencoder}
\acro{ddpm}[DDPM]{denoising diffusion probabilistic model}
\acro{stft}[STFT]{short-time Fourier transform}
\acro{istft}[iSTFT]{inverse short-time Fourier transform}
\acro{sde}[SDE]{stochastic differential equation}
\acro{ode}[ODE]{ordinary differential equation}
\acro{ou}[OU]{Ornstein-Uhlenbeck}
\acro{ve}[VE]{Variance Exploding}
\acro{dnn}[DNN]{deep neural network}
\acro{pesq}[PESQ]{Perceptual Evaluation of Speech Quality}
\acro{se}[SE]{speech enhancement}
\acro{tf}[T-F]{time-frequency}
\acro{elbo}[ELBO]{evidence lower bound}
\acro{WPE}{weighted prediction error}
\acro{PSD}{power spectral density}
\acro{RIR}{room impulse response}
\acro{LSTM}{long short-term memory}
\acro{POLQA}{Perceptual Objectve Listening Quality Analysis}
\acro{SDR}{signal-to-distortion ratio}
\acro{ESTOI}{Extended Short-Term Objective Intelligibility}
\acro{ELR}{early-to-late reverberation ratio}
\acro{TCN}{temporal convolutional network}
\acro{DRR}{direct-to-reverberant ratio}
\acro{nfe}[NFE]{number of function evaluations}
\acro{rtf}[RTF]{real-time factor}
\acro{mos}[MOS]{mean opinion scores}
\acro{asr}[ASR]{automatic speech recognition}
\acro{wer}[WER]{word error rate}
\acro{avse}[AVSE]{audio-visual speech enhancement}
\acro{ssl}[SSL]{self-supervised learning}
\acro{dl}[DL]{deep learning}
\acro{roi}[ROI]{region-of-interest}
\end{acronym}

\maketitle

\begin{abstract}
This paper introduces an audio-visual speech enhancement system that leverages score-based generative models, also known as diffusion models, conditioned on visual information. In particular, we exploit audio-visual embeddings obtained from a self-super\-vised learning model that has been fine-tuned on lipreading. The layer-wise features of its transformer-based encoder are aggregated, time-aligned, and incorporated into the noise conditional score network. Experimental evaluations show that the proposed audio-visual speech enhancement system yields improved speech quality and reduces generative artifacts such as phonetic confusions with respect to the audio-only equivalent. The latter is supported by the word error rate of a downstream automatic speech recognition model, which decreases noticeably, especially at low input signal-to-noise ratios. 
\end{abstract}
\vspace{0.5em}
\vspace{-0.5em}

\nnfootnote{This work has been funded by the German Research Foundation (DFG) in the transregio project Crossmodal Learning (TRR 169). We would like to thank J. Berger and Rohde\&Schwarz SwissQual AG for their support with POLQA.}
\section{Introduction}

Robust speech processing, e.g. in telecommunications or for \ac{asr}, often requires improved speech quality and intelligibility. For this purpose, speech enhancement aims at estimating clean speech signals from noisy audio recordings \cite{gerkmann2018book_chapter}. However, in various real-world scenarios, such as in negative \acp{snr}, speech enhancement systems may face limitations, typically resulting in speech distortions and impaired speech recognition. 

To mitigate these limitations, promising approaches have been proposed that leverage both audio and video information which is referred to as \ac{avse} \cite{michelsanti2021overview}. In fact, incorporating visual information into audio-based systems becomes an emerging research direction and improves both the robustness and the overall performance \cite{zhu2021deep}. The principle is in general twofold: first, the visual information (e.g. lip movements) is usually not affected by the acoustic environment, and second, it has been proven that visual information is able to provide additional speech and speaker-related cues \cite{sumby1954visual, mcgurk1976hearing}. 

Prevailing deep learning approches proposed for \ac{avse} typically involve models with modality-specific encoders, feature fusion, and a decoder, all trained end-to-end in a supervised fashion \cite{hou2018audio, afouras2018conversation}. Their goal is to learn audio-visual representations suitable for speech enhancement which is usually defined as a predictive regression task. Another approach to speech enhancement is to use generative models which follow a different learning paradigm. They aim to model the underlying statistical distribution of clean speech and do not directly estimate the mapping from noisy input to clean output. 

Recently, score-based generative models have been proposed for speech enhancement \cite{welker2022speech, richter2022journal}. The idea is to corrupt clean speech with slowly increasing levels of noise and train a \ac{dnn} to reverse this corruption iteratively. This exciting new research direction provides an alternative to other generative approaches and differs from predictive approaches in its methodology and results. Since the models are trained to generate clean speech conditioned on the noisy input, they tend to hallucinate for signals with low \acp{snr} \cite{richter2022journal}. This can lead to speech-like sounds with poor articulation and no linguistic meaning or cause phonetic confusions, which may be problematic in many practical applications.

\begin{figure}[t]
    \centering
    \includegraphics[scale=0.8]{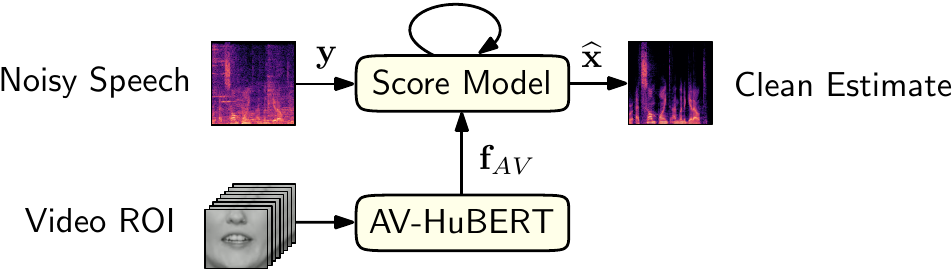}
    \caption{Proposed audio-visual speech enhancement system. The score model iteratively refines noisy speech $\mathbf y$ to obtain a clean estimate $\widehat{\mathbf x}$ and is conditioned on audio-visual embeddings $\mathbf f_{AV}$ obtained from AV-HuBERT.}
    \vspace{-0.7em}
    \label{fig:av_gen_overview}
\end{figure}

In this paper, we build upon our previous work on score-based generative models for speech enhancement \cite{richter2022journal} and develop an audio-visual extension of the generative approach. We condition the noise conditional score network on audio-visual embeddings, that are obtained from AV-HuBERT \cite{shi2022learning}, a  \ac{ssl} model fine-tuned on lipreading (see Fig.~\ref{fig:av_gen_overview}). This follows a recent trend in deep learning-based speech processing to leverage powerful representations from pre-trained models, which have been trained on large-scale datasets using \ac{ssl} techniques. 

Experimental evaluations show that the proposed \ac{avse} system provides better speech quality and reduces generative artifacts, such as phonetic confusions, compared to its audio-only counterpart. This observation is supported by the reduction of the \ac{wer} of a downstream \ac{asr} model, which is particularly evident at low input \acp{snr}.

\vspace{-0.2em}

\section{Related Work}

\subsection{Score-based Generative Models}

Score-based generative models, also known as diffusion models, have recently gained attention due to their ability to generate high-quality samples and provide flexible and efficient training procedures. Originally inspired by non-equilibrium thermodynamics \cite{sohl2015deep}, they exist in several variants \cite{ho2020denoising,song2019generative}. All of them share the idea of gradually turning data into noise, and training a \ac{dnn} that learns to invert this diffusion process for different noise scales. In an iterative refinement procedure, the \ac{dnn} is evaluated multiple times to generate data, achieving a
trade-off between computational complexity and sample quality.

Song et al.~\cite{song2021sde} showed that the diffusion process can be expressed as a solution of a \ac{sde}, establishing a continuous-time formulation for diffusion models. The forward \ac{sde} can be inverted in time, resulting in a corresponding reverse \ac{sde} which depends only on the score function of the perturbed data \cite{anderson1982reverse}.

\subsection{Audio-Visual Speech Enhancement}

Early works on \ac{avse} can be traced back to the 1990s~\cite{girin1995noisy}. More recently, a number of interesting algorithms have been developed in the context of deep learning, including methods based on spectral mapping/masking that are trained in a supervised fashion \cite{hou2018audio, afouras2018conversation}. These methods typically consist of two modality-specific encoders, one for processing noisy speech and another for processing video information. The resulting audio and visual features are then concatenated and processed to obtain an audio-visual embedding that is fed into a decoder to predict the clean speech.

Alternatively, Sadeghi et al.~\cite{sadeghi2020audio} proposed a generative approach to \ac{avse} using a conditional variational autoencoder trained on clean speech only. The model learns a clean speech prior that is conditioned on visual information. At inference, this prior is combined with a noise model based on nonnegative matrix factorization to estimate a probabilistic Wiener filter for denoising.

Other approaches consider \ac{avse} as a speech resynthesis task, i.e., to create an output that closely resembles the original signal in terms of sound
quality, prosody, and other acoustic characteristics \cite{mehrish2023review}. Yang et al.~\cite{yang2022audio} developed a speaker-dependent model that generates clean speech codes of a neural audio codec conditioned on noisy audio and visual inputs. The discrete codes obtained are used to reconstruct a mel-spectrogram which is then fed into a neural vocoder. Hsu et al.~\cite{hsu2023revise} break \ac{avse} into audio-visual speech recognition and text-to-speech exploiting hidden units from a pre-trained  \ac{ssl}-model. Mira et al.~\cite{mira2023voce} proposed a two-stage approach that predicts clean mel-spectrograms from audio-visual speech and converts them into waveform audio using a neural vocoder.

\subsection{Self-Supervised Learning}

Recently, \ac{ssl} has gained significant attention in the speech community, with well-known models such as wav2vec 2.0 \cite{baevski2020wav2vec} and HuBERT \cite{hsu2021hubert} for audio-only and AV-HuBERT \cite{shi2022learning} for audio-visual speech representation learning. In contrast to supervised learning, which involves direct optimization of models for a specific task, \ac{ssl} adopts a two-step approach. First, the model undergoes pre-training on large-scale unlabeled data to extract task-agnostic representations. Then, these learned representations are either used as input to a supervised model, or the model itself is fine-tuned for a particular downstream task. The expected outcome is to either improve the downstream task's performance or reduce the amount of labeled data required for training.

Pasad et al.~\cite{pasad2021layer} have analyzed the layer-wise features of wav2vec 2.0 and found that the transformer layers follow an autoencoder-like behavior in which the representation deviates from the input speech features as the depth of the model increases, followed by a reverse trend in which even deeper layers become more similar to the input. Furthermore, the evolution of the representations follows an acoustic-linguistic hierarchy. The shallowest layers encode acoustic features, followed by phonetic, word identity, and word meaning information \cite{pasad2021layer}. Huang et al.~\cite{huang2022investigating} have investigated \ac{ssl} model for speech enhancement and separation and concluded although they are not designed for waveform generation tasks like enhancement and separation, some of them achieve remarkable improvements over spectrogram magnitudes. Interestingly, instead of extracting representations from the last hidden layer, they proposed to aggregate embeddings from all layers with a learnable weighted sum. 

\section{Method}

In this section, we describe the proposed \ac{avse} system, which we name \emph{AV-Gen}. Fig.~\ref{fig:av_gen_overview} shows an overview of AV-Gen which consists of a score model and a pre-trained AV-HuBERT model as an upstream video processor.

\subsection{Score Model}

Following the work in \cite{richter2022journal}, we cast speech enhancement as a conditional generation task using score-based generative models. 
For a pair of clean and noisy spectrograms $\mathbf x_0, 
\mathbf y \in \mathbb C^d$ with $d$ denoting the number of time-frequency bins, we define a stochastic diffusion process $\{\mathbf x_t\}_{t=0}^T$, called \emph{forward process}, as the solution to the \ac{sde}, 
\begin{equation} \label{eq:ouve-sde}
    \mathrm{d}{\mathbf x_t} = \gamma(\mathbf y-\mathbf x_t) \mathrm{d} t + g(t) \mathrm{d} \mathbf w\,,
\end{equation}
where $\mathbf x_t$ denotes the process state at time $t \in [0, T]$, $\gamma \in \mathbb R$  controls the transition from $\mathbf x_0$ to $\mathbf y$, and $g(t) \in \mathbb R$ is the diffusion coefficient that controls the amount of Gaussian white noise induced by a standard Wiener process $\mathbf w$.

The forward process has an associated \emph{reverse process} which is given by the reverse \ac{sde} \cite{anderson1982reverse,song2021sde},
\begin{equation}\label{eq:reverse-sde}
\begin{split}
\mathrm{d} \mathbf x_t =
        [
            -\gamma(\mathbf y-\mathbf x_t) + g(t)^2 \underbrace{\nabla_{\mathbf x_t} \log p_t(\mathbf x_t|\mathbf y)}_{\approx \,\mathbf s_\theta(\mathbf x_t, \mathbf y, \mathbf f_{AV}, t)}
        ] \mathrm{d}t + g(t)\mathrm{d} \bar{\mathbf w}\,,
\end{split}
\end{equation}
where the score function $\nabla_{\mathbf x_t} \log p_t(\mathbf x_t|\mathbf y)$ is intractable and therefore approximated by the score model $\mathbf s_\theta$ with learnable parameters $\theta$. As in \cite{richter2022journal}, we use the Noise Conditional Score Network (NCSN++) \cite{song2021sde} as the score model. The training objective is denoising score matching, as defined in \cite[Eq.~(15)]{richter2022journal}. Note, however, that in contrast to \cite{richter2022journal}, the score model is additionally conditioned on the video modality by taking conditional features $\mathbf f_{AV}$ as input.

At inference, we can generate a clean speech estimate $\widehat{\mathbf x}$ by solving the reverse \ac{sde} in Eq.~\eqref{eq:reverse-sde}. For this purpose, a numerical \ac{sde} solver is employed which solves the task of speech enhancement iterating through the reverse process starting at $t=T$ and ending at $t=t_\epsilon$, where $t_\epsilon$ is a small minimum process time to avoid numerical instabilities.

\subsection{AV-HuBERT as Conditional Model}

\begin{figure}[t]
    \centering
    \includegraphics[scale=0.8]{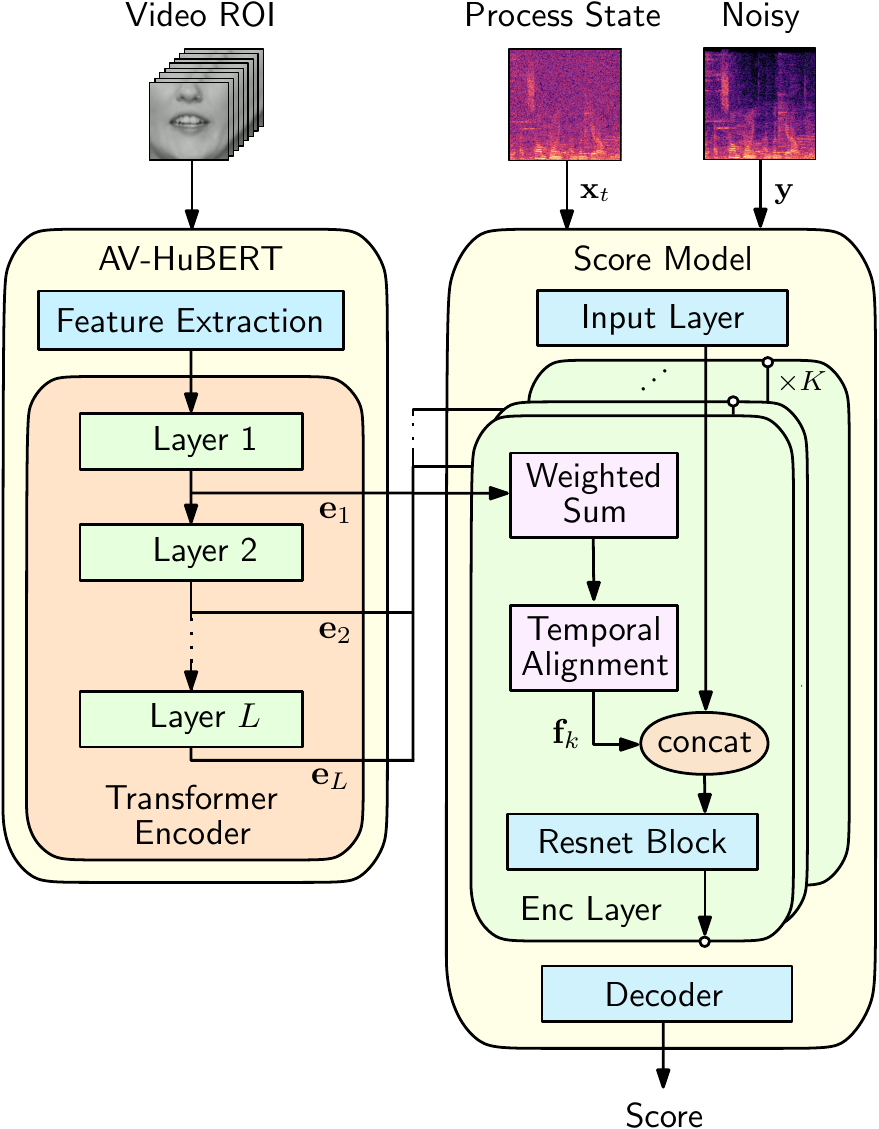}
    \caption{Network architecture: Each encoder layer $k$ of the score model is conditioned on a conditional feature $\mathbf f_k$ which is a weighted sum of the layerwise embedding $\mathbf e_l$ of the AV-HuBERT's transformer encoder. }
    \label{fig:feature_fusion}
\end{figure}

As an upstream video processor for the score model, we use a pre-trained AV-HuBERT model consisting of a hybrid ResNet-transformer architecture for which we freeze the parameters at training. In particular, we exploit
audio-visual embeddings from each layer of its transformer encoder, as visualized in Fig. \ref{fig:feature_fusion}. For each encoder layer $k$ of the score model, we learn a conditional feature $\mathbf f_k$ by aggregating the layer-wise embeddings $\mathbf e_l$ of AV-HuBERT's transformer encoder, as
\begin{equation}
    \mathbf f_k = \operatorname{TempAlign}_k \left( \sum_{l=1}^{L} w_{kl} \, \mathbf e_l \right), \quad k = 1, \dots, K\,,
\end{equation}
with trainable weights $w_{kl}$ with the properties $w_{kl}\geq 0$ and $\sum_l w_{kl} = 1$ for all $k$. The operation $\operatorname{TempAlign}_k(\cdot)$  ensures that the features are temporally aligned to the feature rates of the score model's embeddings for different resolutions. In particular, we use linear interpolation for upsampling and trainable Conv1D layers with kernel size and stride $d$ for downsampling with factor $d$. The features are broadcasted over the frequency dimension to match the feature maps of the score model and the number of channels is adopted in the temporal alignment operation. Finally, the conditional features $\mathbf f_{AV} = \{\mathbf f_1, \dots, \mathbf f_K\}$ have the same shapes as the embeddings of the score model before concatenation.

\section{Experimental Setup}

\subsection{Dataset}

For audio-visual speech data, we use the LRS3 dataset \cite{afouras2018lrs3}, a large-scale dataset primarily designed for the task of audio-visual speech recognition. Besides the audio tracks sampled at $16\,\text{kHz}$, it includes face tracks sampled at $25\,\text{Hz}$ from over 400 hours of TED and TEDx videos. Note that recording conditions vary among the files and the data sometimes contain reverberant speech and is not as clean as data recorded in controlled environments such as a recording studio. For practical reasons, we limit the size of training data and use the low-resource 30 hours subset introduced in \cite{shi2022learning}. To meet the requirements for evaluating speech quality \cite{rixPerceptualEvaluationSpeech2001}, we only consider files from the test set with a minimum length of $3.2\,\text{s}$ resulting in 244 test files in total.

We generate pairs of clean and noisy speech by mixing the audio tracks of LRS3 with noise signals from the CHiME3 dataset \cite{barker2015third} at \acp{snr} sampled uniformly between -6 and 12$\,$dB for the training, validation, and test. Consequently, we name the resulting data set LRS3-CHiME3.

\subsection{Baselines}

The primary objective of this paper is to investigate the impact of incorporating visual information into SGMSE+ \cite{richter2022journal}, our previous work on score-based generative models for speech enhancement. To accomplish this, we train SGMSE+ using identical settings as those employed for AV-Gen. This encompasses the input representation, \ac{sde} parameters, and matching training configurations, including batch size and learning rate. Considering that both methods employ the same \ac{dnn} as the score model, with the only distinction being the visual conditioning, it is reasonable to regard SGMSE+ as the audio-only counterpart to AV-Gen.

Furthermore, we consider VisualVoice \cite{gao2021visualvoice} as an audio-visual baseline, a  method originally proposed for audio-visual speech separation. This method learns a direct mapping from noisy to clean speech conditioned on visual information. To this end, it utilizes an encoder-decoder architecture (54,7M parameters) for complex mask prediction, in which the audio features are combined with visual features extracted from a lip motion analysis network. Here, we only use the mask prediction loss and omit the facial analysis network, as we adopt the method for \ac{avse}.

\begin{table*}[t]
\centering
\resizebox{\textwidth}{!}{
\begin{tabular}{l|c|cccc|cccc}
\toprule
Method & AV & POLQA & PESQ & ESTOI & SI-SDR [dB] & WER & S & D & I \\
\midrule
Mixture &   & $1.68 \pm 0.55$ & $1.20 \pm 0.20$ & $0.77 \pm 0.13$ & $\:\:2.6 \pm 5.1$ & $0.28 \pm 0.30$ & $2.2$ & $1.5$ & $\mathbf{0.1}$ \\
SGMSE+ \cite{richter2022journal} &   & $2.73 \pm 0.61$ & $2.08 \pm 0.56$ & $0.89 \pm 0.08$ & $10.6 \pm 4.8$ & $0.22 \pm 0.25$ & $2.3$ & $0.6$ & $0.2$ \\
VisualVoice \cite{gao2021visualvoice} & \checkmark & $2.78 \pm 0.60$ & $2.11 \pm 0.57$ & $0.89 \pm 0.07$ & $11.0 \pm 4.0$ & $0.15 \pm 0.18$ & $1.5$ & $\mathbf{0.3}$ & $0.2$ \\
AV-Gen (ours) & \checkmark & $\mathbf{2.87 \pm 0.55}$ & $\mathbf{2.23 \pm 0.53}$ & $\mathbf{0.90 \pm 0.06}$ & $\mathbf{11.4 \pm 3.9}$ & $\mathbf{0.11 \pm 0.14}$ & $\mathbf{1.1}$ & $\mathbf{0.3}$ & $0.2$ \\
\bottomrule
\end{tabular}
}
\caption{Average speech enhancement performance obtained for the LRS3+CHiME3 test set. Values indicate mean and standard deviation for the 244 test files.}
\label{tab:results}
\end{table*}

\subsection{Hyperparameter Setting}

\subsubsection{Input Representation}

The complex spectrogram is computed with the \ac{stft} using a Hann window of length 510 and a hop size of length 160 resulting in 256
unique frequency bins and an audio frame rate of $100\,\text{Hz}$. Consequently, the audio frame rate is a multiple of the video frame rate, which provides simple temporal alignment for feature fusion to ensure audio-visual synchronization. We use the same spectrogram transformation as in \cite{richter2022journal}. Unlike the original work, we do not normalize the signal in the time domain, which should make the score model more robust to varying input conditions.

For video pre-processing, we follow the protocol in \cite{shi2022learning} which stems from prior works in lipreading \cite{martinez2020lipreading, ma2021end}. However, instead of using \texttt{dlib} for facial landmark detection, we use the Face Alignment Network \cite{bulat2017far} because we found better robustness when inspecting the processed data. Using the 68 facial landmarks, each video clip is aligned to a reference face frame via affine transformation and cropped to a $88 \times 88$ \ac{roi} centered on the mouth. 

\subsubsection{Score model and training configuration}

We parameterize the \ac{sde} in Eq.~\eqref{eq:ouve-sde} and the NCSN++ architecture with the same configuration as in \cite{richter2022journal}, resulting in 76M learnable parameters.  We train the score model with variable length sequences (between $2$ and $12$ seconds), using the \texttt{MaxTokenBucketizer} from TorchData, which forms batches of variable batch size. We utilize four Quadro RTX 6000 (24 GB memory each) and train for 60 epochs using the distributed data-parallel (DDP) approach in \mbox{PyTorch} Lightning, which takes about $2.5$ days. We use the Adam optimizer~\cite{kingma2015adam} with a learning rate of $5\cdot 10^{-4}$.

\subsubsection{AV-HuBERT}

We use AV-HuBERT Base with $L=12$ transformer layers. We use the fine-tuned model for visual speech recognition that used LRS3 as pretraining data and LRS3-433h as finetuning data. This model takes only the video modality as input. We also experimented with the pre-trained model trained on LRS3 using both modalities as input, which resulted in inferior performance.

\subsection{Metrics}

For an instrumental evaluation of speech quality and intelligibility, we use standard metrics, including POLQA \cite{polqa2018}, PESQ \cite{rixPerceptualEvaluationSpeech2001}, ESTOI \cite{jensen2016algorithm}, and SI-SDR \cite{leroux2018sdr}.

To evaluate the effect of speech enhancement on \ac{asr}, we use QuartzNet (Quartz\-Net\-15\-x5Base-En) from the NeMo toolkit \cite{kuchaiev2019nemo} as a downstream \ac{asr} system. We measure the \ac{wer}, which is defined as
\begin{equation}
    \operatorname{WER} = \frac{S+D+I}{N}\, ,
\end{equation}
where $S$ is the number of substitutions, $D$ is the number of deletions, $I$ is the number of insertions needed to get from the predicted to the target sequence, and $N$ is the number of words in the target sequence.

\section{Results}

In Tab. \ref{tab:results}, we report the speech enhancement performance obtained for the LRS3-CHiME3. Comparing AV-Gen to its audio-only counterpart SGMSE+, we see that the inclusion of video information improves the speech quality, indicated by an improvement of $0.14$ in POLQA and $0.15$ in PESQ. \mbox{ESTOI} is only marginally better, suggesting that both methods are fairly similar in terms of intelligibility. However, the evaluation on \ac{asr} shows an improved \ac{wer} of 0.11 on average. We argue that this is mainly because SGMSE+ tends to produce generative artifacts like phonetic confusions at low input \acp{snr} \cite{richter2022journal}, which is alleviated by using video information. In fact, the reduced number of substitutions and deletions on average reveals a stronger robustness with respect to generative hallucinations, which is also verified by informal listening. Moreover, AV-Gen seems to be competitive with the 
 audio-visual baseline VisualVoice, outperforming it in all metrics.  

\begin{figure}[t]
    \centering
    \includegraphics[scale=0.75]{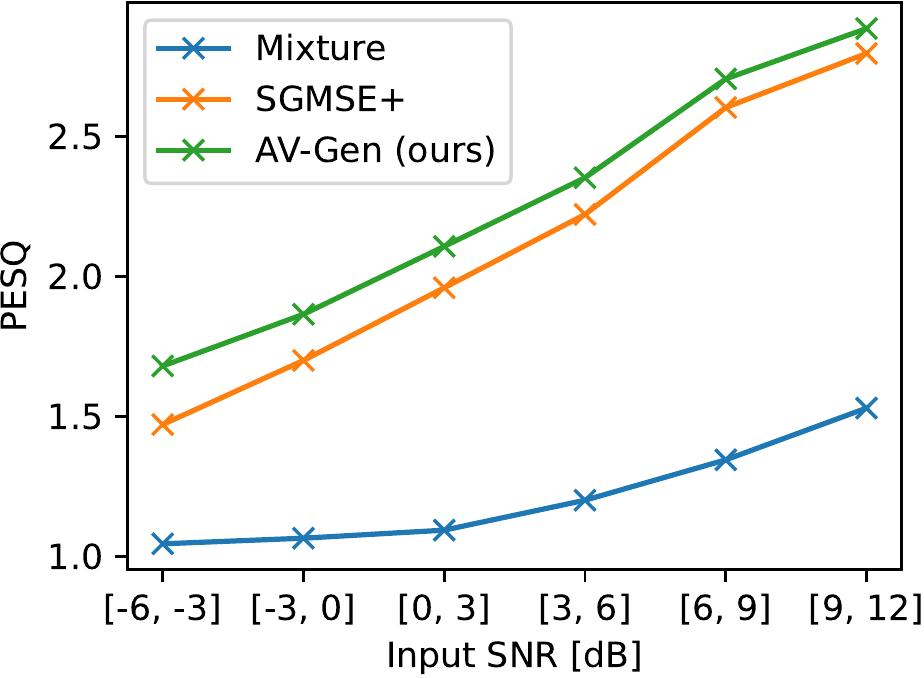}
    \caption{Mean performance in PESQ evaluated for different ranges of input \acp{snr}.}
    \label{fig:pesq_per_snr}
\end{figure}

\begin{figure}[t]
    \centering
    \includegraphics[scale=0.75]{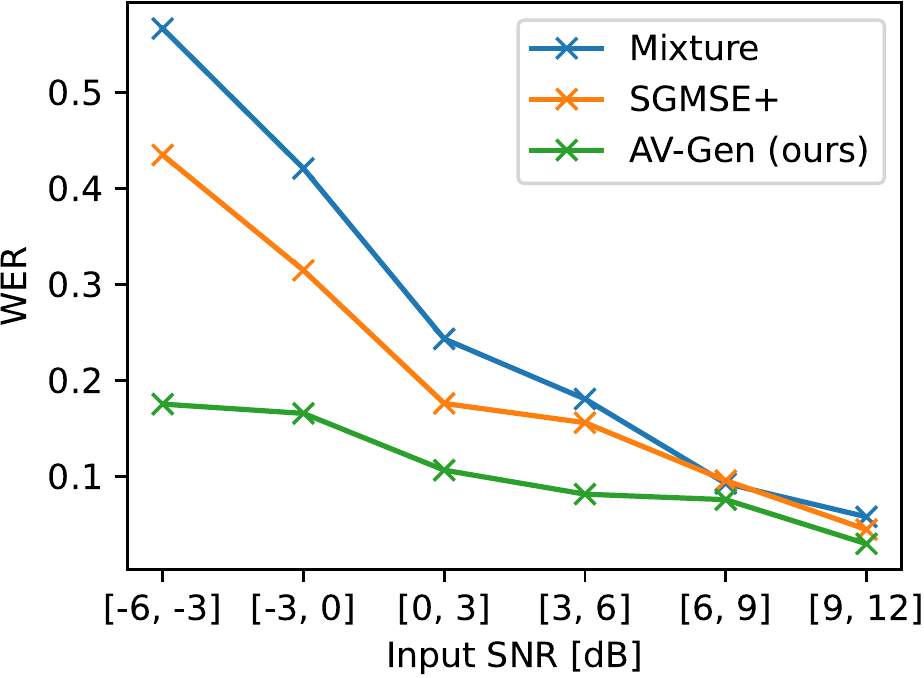}
    \caption{Mean performance in WER evaluated for different ranges of input \acp{snr}}
    \label{fig:wer_per_snr}
\end{figure}

In informal listening, we find that AV-Gen occasionally generates a more reverberant output compared to the noisy input. This could be due to the generative behavior of the proposed model, as the data used for training was recorded in the wild and also contains reverberant speech. 
To prevent this effect, the training data could be cleaned beforehand.

To gain more insights into the effect of exploiting video information in AV-Gen as opposed to SGMSE+, we evaluate PSEQ and \ac{wer} at different input \acp{snr}. Fig. \ref{fig:pesq_per_snr} shows the mean performance in PESQ, and Fig. \ref{fig:wer_per_snr} for WER, when evaluated at specific ranges of input \acp{snr}. As expected, both metrics degrade for decreasing input \acp{snr}. However, the proposed audio-visual method shows stronger robustness, indicated by larger relative improvements. This is particularly evident for \ac{wer}, with drastic degradations for SGMSE+ at negative input \acp{snr}, at which AV-Gen remains fairly robust. Please find some listening examples on our project page\footnote{\url{https://uhh.de/inf-sp-av-gen}}.

\section{Conclusion}

In this paper, we presented an \ac{avse} system that utilizes score-based generative models conditioned on visual information. To this end, we leverage audio-visual embeddings obtained from AV-HuBERT, a self-supervised learning model that was fine-tuned on lipreading. The experimental evaluations demonstrated that the proposed method improves speech quality and noticeably reduces generative artifacts, such as phonetic confusions, in comparison to its audio-only counterpart. The effectiveness of the system is further supported by the significant reduction in \ac{wer}, as observed in a downstream \ac{asr} model, particularly for low input \acp{snr}.

\clearpage

\small
\bibliographystyle{ieeetr}
\bibliography{bib_clean}

\begin{thebibliography}{10}

\bibitem{gerkmann2018book_chapter}
T.~Gerkmann and E.~Vincent, ``Spectral masking and filtering,'' in {\em Audio
  Source Separation and Speech Enhancement} (E.~Vincent, T.~Virtanen, and
  S.~Gannot, eds.), John Wiley \& Sons, 2018.

\bibitem{michelsanti2021overview}
D.~Michelsanti, Z.-H. Tan, S.-X. Zhang, Y.~Xu, M.~Yu, D.~Yu, and J.~Jensen,
  ``An overview of deep-learning-based audio-visual speech enhancement and
  separation,'' {\em IEEE Trans. on Audio, Speech, and Language Proc. (TASLP)},
  vol.~29, pp.~1368--1396, 2021.

\bibitem{zhu2021deep}
H.~Zhu, M.-D. Luo, R.~Wang, A.-H. Zheng, and R.~He, ``Deep audio-visual
  learning: A survey,'' {\em Int. Journal of Automation and Computing},
  vol.~18, pp.~351--376, 2021.

\bibitem{sumby1954visual}
W.~H. Sumby and I.~Pollack, ``Visual contribution to speech intelligibility in
  noise,'' {\em The Journal of the Acoustical Society of America}, vol.~26,
  no.~2, pp.~212--215, 1954.

\bibitem{mcgurk1976hearing}
H.~McGurk and J.~MacDonald, ``Hearing lips and seeing voices,'' {\em Nature},
  vol.~264, no.~5588, pp.~746--748, 1976.

\bibitem{hou2018audio}
J.-C. Hou, S.-S. Wang, Y.-H. Lai, Y.~Tsao, H.-W. Chang, and H.-M. Wang,
  ``Audio-visual speech enhancement using multimodal deep convolutional neural
  networks,'' {\em IEEE Transactions on Emerging Topics in Computational
  Intelligence}, vol.~2, no.~2, pp.~117--128, 2018.

\bibitem{afouras2018conversation}
T.~Afouras, J.~S. Chung, and A.~Zisserman, ``The conversation: Deep
  audio-visual speech enhancement,'' {\em Proc. Interspeech 2018},
  pp.~3244--3248, 2018.

\bibitem{welker2022speech}
S.~Welker, J.~Richter, and T.~Gerkmann, ``Speech enhancement with score-based
  generative models in the complex {STFT} domain,'' {\em ISCA Interspeech},
  pp.~2928--2932, 2022.

\bibitem{richter2022journal}
J.~Richter, S.~Welker, J.-M. Lemercier, B.~Lay, and T.~Gerkmann, ``Speech
  enhancement and dereverberation with diffusion-based generative models,''
  {\em IEEE Trans. on Audio, Speech, and Language Proc. (TASLP)}, 2023.

\bibitem{shi2022learning}
B.~Shi, W.-N. Hsu, K.~Lakhotia, and A.~Mohamed, ``Learning audio-visual speech
  representation by masked multimodal cluster prediction,'' in {\em Int. Conf.
  on Learning Repr. (ICLR)}, 2022.

\bibitem{sohl2015deep}
J.~Sohl-Dickstein, E.~Weiss, N.~Maheswaranathan, and S.~Ganguli, ``Deep
  unsupervised learning using nonequilibrium thermodynamics,'' {\em Int. Conf.
  on Machine Learning (ICML)}, pp.~2256--2265, 2015.

\bibitem{ho2020denoising}
J.~Ho, A.~Jain, and P.~Abbeel, ``Denoising diffusion probabilistic models,''
  {\em Advances in Neural Inf. Proc. Systems (NeurIPS)}, vol.~33,
  pp.~6840--6851, 2020.

\bibitem{song2019generative}
Y.~Song and S.~Ermon, ``Generative modeling by estimating gradients of the data
  distribution,'' {\em Advances in Neural Inf. Proc. Systems (NeurIPS)},
  vol.~32, 2019.

\bibitem{song2021sde}
Y.~Song, J.~Sohl-Dickstein, D.~P. Kingma, A.~Kumar, S.~Ermon, and B.~Poole,
  ``Score-based generative modeling through stochastic differential
  equations,'' {\em Int. Conf. on Learning Repr. (ICLR)}, 2021.

\bibitem{anderson1982reverse}
B.~D. Anderson, ``Reverse-time diffusion equation models,'' {\em Stochastic
  Processes and their Applications}, vol.~12, no.~3, pp.~313--326, 1982.

\bibitem{girin1995noisy}
L.~Girin, G.~Feng, and J.-L. Schwartz, ``Noisy speech enhancement with filters
  estimated from the speaker's lips.,'' in {\em Eurospeech}, 1995.

\bibitem{sadeghi2020audio}
M.~Sadeghi, S.~Leglaive, X.~Alameda-Pineda, L.~Girin, and R.~Horaud,
  ``Audio-visual speech enhancement using conditional variational
  auto-encoders,'' {\em IEEE Trans. on Audio, Speech, and Language Proc.
  (TASLP)}, vol.~28, pp.~1788--1800, 2020.

\bibitem{mehrish2023review}
A.~Mehrish, N.~Majumder, R.~Bhardwaj, and S.~Poria, ``A review of deep learning
  techniques for speech processing,'' {\em arXiv preprint arXiv:2305.00359},
  2023.

\bibitem{yang2022audio}
K.~Yang, D.~Markovi{\'c}, S.~Krenn, V.~Agrawal, and A.~Richard, ``Audio-visual
  speech codecs: Rethinking audio-visual speech enhancement by re-synthesis,''
  in {\em Proceedings of the IEEE/CVF Conference on Computer Vision and Pattern
  Recognition}, pp.~8227--8237, 2022.

\bibitem{hsu2023revise}
W.-N. Hsu, T.~Remez, B.~Shi, J.~Donley, and Y.~Adi, ``\mbox{{ReVISE}}:
  Self-supervised speech resynthesis with visual input for universal and
  generalized speech regeneration,'' in {\em Proceedings of the IEEE/CVF
  Conference on Computer Vision and Pattern Recognition}, pp.~18795--18805,
  2023.

\bibitem{mira2023voce}
R.~Mira, B.~Xu, J.~Donley, A.~Kumar, S.~Petridis, V.~K. Ithapu, and M.~Pantic,
  ``{LA-VocE}: Low-{SNR} audio-visual speech enhancement using neural
  vocoders,'' in {\em IEEE International Conference on Acoustics, Speech and
  Signal Processing (ICASSP)}, pp.~1--5, 2023.

\bibitem{baevski2020wav2vec}
A.~Baevski, Y.~Zhou, A.~Mohamed, and M.~Auli, ``wav2vec 2.0: A framework for
  self-supervised learning of speech representations,'' {\em Advances in Neural
  Inf. Proc. Systems (NeurIPS)}, vol.~33, 2020.

\bibitem{hsu2021hubert}
W.-N. Hsu, B.~Bolte, Y.-H.~H. Tsai, K.~Lakhotia, R.~Salakhutdinov, and
  A.~Mohamed, ``Hubert: Self-supervised speech representation learning by
  masked prediction of hidden units,'' {\em IEEE Trans. on Audio, Speech, and
  Language Proc. (TASLP)}, vol.~29, pp.~3451--3460, 2021.

\bibitem{pasad2021layer}
A.~Pasad, J.-C. Chou, and K.~Livescu, ``Layer-wise analysis of a
  self-supervised speech representation model,'' in {\em 2021 IEEE Automatic
  Speech Recognition and Understanding Workshop (ASRU)}, pp.~914--921, IEEE,
  2021.

\bibitem{huang2022investigating}
Z.~Huang, S.~Watanabe, S.-w. Yang, P.~Garc{\'\i}a, and S.~Khudanpur,
  ``Investigating self-supervised learning for speech enhancement and
  separation,'' in {\em IEEE Int. Conf. on Acoustics, Speech and Signal Proc.
  (ICASSP)}, pp.~6837--6841, 2022.

\bibitem{afouras2018lrs3}
T.~Afouras, J.~S. Chung, and A.~Zisserman, ``Lrs3-ted: a large-scale dataset
  for visual speech recognition,'' {\em arXiv preprint arXiv:1809.00496}, 2018.

\bibitem{rixPerceptualEvaluationSpeech2001}
A.~Rix, J.~Beerends, M.~Hollier, and A.~Hekstra, ``Perceptual evaluation of
  speech quality ({{PESQ}}) - a new method for speech quality assessment of
  telephone networks and codecs,'' {\em IEEE Int. Conf. on Acoustics, Speech
  and Signal Proc. (ICASSP)}, vol.~2, pp.~749--752, 2001.

\bibitem{barker2015third}
J.~Barker, R.~Marxer, E.~Vincent, and S.~Watanabe, ``The third '{CHiME}' speech
  separation and recognition challenge: Dataset, task and baselines,'' {\em
  IEEE Workshop on Automatic Speech Recognition and Understanding (ASRU)},
  pp.~504--511, 2015.

\bibitem{gao2021visualvoice}
R.~Gao and K.~Grauman, ``Visualvoice: Audio-visual speech separation with
  cross-modal consistency,'' in {\em 2021 IEEE/CVF Conference on Computer
  Vision and Pattern Recognition (CVPR)}, pp.~15490--15500, IEEE, 2021.

\bibitem{martinez2020lipreading}
B.~Martinez, P.~Ma, S.~Petridis, and M.~Pantic, ``Lipreading using temporal
  convolutional networks,'' in {\em IEEE International Conference on Acoustics,
  Speech and Signal Processing (ICASSP)}, pp.~6319--6323, 2020.

\bibitem{ma2021end}
P.~Ma, S.~Petridis, and M.~Pantic, ``End-to-end audio-visual speech recognition
  with conformers,'' in {\em IEEE International Conference on Acoustics, Speech
  and Signal Processing (ICASSP)}, pp.~7613--7617, 2021.

\bibitem{bulat2017far}
A.~Bulat and G.~Tzimiropoulos, ``How far are we from solving the {2D} \& {3D}
  face alignment problem? (and a dataset of 230,000 {3D} facial landmarks),''
  in {\em International Conference on Computer Vision}, 2017.

\bibitem{kingma2015adam}
D.~P. Kingma and J.~Ba, ``Adam: A method for stochastic optimization,'' {\em
  Int. Conf. on Learning Repr. (ICLR)}, 2015.

\bibitem{polqa2018}
{ITU-T Rec. P.863}, ``Perceptual objective listening quality prediction,'' {\em
  Int. Telecom. Union ({ITU})}, 2018.

\bibitem{jensen2016algorithm}
J.~Jensen and C.~H. Taal, ``An algorithm for predicting the intelligibility of
  speech masked by modulated noise maskers,'' {\em IEEE Trans. on Audio,
  Speech, and Language Proc. (TASLP)}, vol.~24, no.~11, pp.~2009--2022, 2016.

\bibitem{leroux2018sdr}
J.~Le~Roux, S.~Wisdom, H.~Erdogan, and J.~R. Hershey, ``{SDR}--half-baked or
  well done?,'' {\em IEEE Int. Conf. on Acoustics, Speech and Signal Proc.
  (ICASSP)}, pp.~626--630, 2019.

\bibitem{kuchaiev2019nemo}
O.~Kuchaiev, J.~Li, H.~Nguyen, O.~Hrinchuk, R.~Leary, B.~Ginsburg, S.~Kriman,
  S.~Beliaev, V.~Lavrukhin, J.~Cook, {\em et~al.}, ``{NeMo}: a toolkit for
  building {AI} applications using neural modules,'' {\em arXiv preprint
  arXiv:1909.09577}, 2019.

\end{thebibliography}


\end{document}